\documentclass[runningheads]{llncs}
\usepackage{graphicx,booktabs}
\usepackage{amssymb,xcolor}
\usepackage{amsmath}

\begin{document}

\title{Reconstructing the Hemodynamic Response Function via a Bimodal Transformer}
\titlerunning{Hemodynamic Response Function Transformer}
%
\author{Yoni Choukroun$^1$, Lior Golgher$^2$, Pablo Blinder$^{3,4}$, Lior Wolf$^1$}
\authorrunning{Yoni Choukroun et al.}
%
\institute{$^1$The School of Computer Science, Tel Aviv University $^2$ The Edmond and Lily Safra Center for Brain Sciences, The Hebrew University of Jerusalem $^3$Neurobiology, Biochemistry and Biophysics School, Wise Life Science Faculty, Tel Aviv University $^4$The Sagol School for Neuroscience, Tel Aviv University 
}
\maketitle              
\begin{abstract}

The relationship between blood flow and neuronal activity is widely recognized, with blood flow frequently serving as a surrogate for neuronal activity in fMRI studies. At the microscopic level, neuronal activity has been shown to influence blood flow in nearby blood vessels. This study introduces the first predictive model that addresses this issue directly at the explicit neuronal population level. Using in vivo recordings in awake mice, we employ a novel spatiotemporal bimodal transformer architecture to infer current blood flow based on both historical blood flow and ongoing spontaneous neuronal activity. Our findings indicate that incorporating neuronal activity significantly enhances the model's ability to predict blood flow values. Through analysis of the model's behavior, we propose hypotheses regarding the largely unexplored nature of the hemodynamic response to neuronal activity.

\end{abstract}
\section{Introduction}

The brain consumes copious amounts of energy to sustain its activity, resulting in a skewed energetic budget per mass compared to the rest of the body (about 25\%  utilized by about 3\%, see \cite{levy2021communication,buxton2023thermodynamic} for an elaborate review of energy utilization). Given this disproportionate need, resources are allocated on a need-basis: active areas signal to the nearby blood vessel to dilate and increase blood flow, bringing a surplus of resources, to that area. This fundamental physiological process is called neurovascular coupling. It is non-trivial to model and different types of neuronal activity have been shown to elicit opposite vascular responses. 

Neurovascular coupling is a cornerstone of proper brain function and also underpins the ability to observe and study the human brain in action. Imaging methods based on blood oxygenated level dependent (BOLD) approaches rely on it~\cite{Kim2012}, as do methods that are based on rheological properties, such as blood volume and flow speed. Since these methods do not directly measure neuronal activity per-se, but a physiological proxy, i.e. the resulting change in vascular dynamics and oxygen levels, it is of utmost importance to know the precise transform function linking neuronal activity to the observed vascular dynamics. Given the differential response to neuronal activity (see \cite{Drew2019} for a timely review), obtaining a cellular and population level hemodynamic response function (HRF) remains an unmet need in this field, that would finally unlock the ability to infer neuronal activity directly from blood flow dynamics \cite{Logothetis2008a}.

The initial characterization of the hemodynamic response function (HRF) was performed at the \textit{system} level, where \textit{system} refers to large cortical regions encompassing tens of thousands of neurons of different types, without taking into account the fine details of different vascular compartments  (see \cite{Uluda2017} for a succinct review on the original works). At this level, a canonical response function was derived from extensive work on sensory-evoked somatosensory responses. This HRF has become widely accepted and used in the interpretation of BOLD signals. This function consists of three components: an initial dip (its existence and physiological origin are much debated), a prolonged and very pronounced overshoot, followed by a shallower and much shorter undershoot. The initial dip occurs within one second of the sensory stimulus, the overshoot peaks around five seconds later, overshoot and return to baseline level occurs within 15-20 seconds post stimuli. It should be noted that vascular reactivity is much faster than the collective behavior described by the canonical HRF, with reports showing sensory-evoked vascular responses observed after just 300ms. Recently, more advanced imaging and analysis methods have pushed the formulation of an HRF at the single cell to single blood vessel (capillary) level, pointing to a rather narrow family of possible functions. Importantly, this work also established that the HRF derived at the microscopic level can be partially translated to macroscopic imaging approaches. Nevertheless, single neuron to single vessel responses fail to capture the more complex and varied neuronal population level responses that could be integrated across the extensive vascular network that surrounds them. Here, we exploit a unique dataset, in which neuronal and vascular responses (changes in diameter) were recorded in a volumetric fashion and with relevant temporal resolution, allowing us to establish a novel pipeline to uncover/formulate a many-to-many HRF. 

Our model needs to combine neuron firing and blood vessel data and employs a multi-modal transformer. There are three types of multi-modal transformers: (i) a multi-modal Transformer where the two modalities are concatenated and separated by the \texttt{[SEP]} token~\cite{li2019visualbert,li2020oscar}, and self-attention is used, (ii) co-attention-based model modules that contextualize each modality with the other modality~\cite{tan2019lxmert,lu2019vilbert}, and (iii) generative models containing an encoder that uses self-attention on the input and a decoder that uses both the encoded data and data from the decoder's domain as inputs~\cite{carion2020end,zhu2021deformable,wang2020end,paul2021local,wang2020max,vaswani2017attention,lewis2019bart}. Our model is of the third type and presents two distinctive properties: pulling from multiple time points and an attention mechanism that is modulated based on distance.

Our results show that the new transformer model can predict the state of blood vessels better than the baseline models. The utility of neuronal data in the prediction is demonstrated by an ablation study. By analyzing the learned model, we  obtain insights into the link between neuronal and vascular activities.

\section{Data}
All procedures were approved by the Anonymous Ethics Committee for Animal Use and Welfare and followed pertinent Institutional Animal Care and Use Committee (IACUC) and local guidelines.
Neuronal activity was monitored in female C57BL/6J transgenic mice expressing Thy1-GCaMP6s. Vascular dynamics were tracked using a Texas Red fluorescent dye, which was conjugated to a large polysaccharide moiety (2 mega Dalton dextran) and retro-orbitally bolus injected under brief isoflurane sedation at the beginning of the imaging day.

425 quasi-linear vascular segments and 50 putative neuronal cell bodies were manually labeled within a volume of $490\times500\times300{\mu m}^3$, which was continuously imaged across two consecutive 1850-second long sessions at an imaging rate of 30.03 volumes per second. For neuronal activity estimation, we selected a cuboid volume of interest around each neuronal cell body and summed the fluorescence within it following an axial intensity normalization corresponding to an uneven duty cycle of our varifocal lens.

For vascular diameter estimation we used the Radon transform,~\cite{gao2014determination,mookiah2020review,drew2010rapid,asl2017tracking,pourreza2008radon,fazlollahi2014efficient,tavakoli2011radon} as its resilience to rotation and poor contrast are particularly useful for our application. Specifically, Gao and Drew have formerly found that thresholding the vascular intensity profile in Radon space is more resilient to noise than other thresholding methods~\cite{gao2014determination}. Based on their observation, we used the time-collapsed imagery to determine a threshold in Radon space, which was then applied separately for each frame in time. 

This unique ability to rapidly track neuronal and vascular interactions across a continuous brain volume bears several important advantages. 
In particular, a greater proportion of the vascular ensemble that reacts to a given neuronal metabolic demand can be accounted for.

\section{Method}
The HRF learning problem explored in this work is defined as the prediction of current blood flow rates at different vessel segments, given the previous neuronal spikes as well as previous blood flow rates.
We propose to design a parameterized deep neural network $f_{\theta}$ for scalar regression of blood flow rates at different vessel segments, such that at a given time $t$ we have 
\begin{equation}
\begin{aligned}
\label{eq:formal_def}
f_{\theta}(S_{t},F_{t}, X_{S},X_{F})\rightarrow \mathbb{R}^{m}
\end{aligned}
\end{equation}
where the matrix $S_{t}\in \mathbb{R}^{t_{s} \times n}$ denotes the  $n$ neurons' spikes at the $t_{s}$ previous samples, while the matrix $F_{t}\in \mathbb{R}^{t_{v} \times m}$ denotes the blood flow of the $m$ vessel segment at the previous $t_{v}$ time samples. 
$X_{S} \in \mathbb{R}^{n\times3}$ and $X_{F}\in \mathbb{R}^{m\times 3}$ are the three-dimensional positions of the neurons and vessel segments, respectively.

HRF predictions should satisfy fundamental symmetries and invariance of physiological priors and of experimental bias, such as invariance to rigid spatial transformation (rotation and translation).
Therefore, a positional input $X_{u}$ is transformed to inter-elements Euclidean distances $D_{u}=\{d^{u}_{ij}\}_{i,j}$ where \mbox{$d^{u}_{ij}=\|(X_{u})_{i}-(X_{u})_{j}\|_2$} for rigid transform invariance.

Thus, the learning problem is refined as $f_{\theta}: \{ S_{t},F_{t}, D_{S}, D_{F}, D_{SF}\} \rightarrow \mathbb{R}^{m}$,
where $D_{S}, D_{F}, D_{SF}$ represents the Euclidean distance matrix between neurons, vessel segments, and neurons to vessel segments, respectively.
We do not include any further auxiliary features or prior in the input.

We model $f_{\theta}$ using a new variant of the Transformer family. 
The proposed model consists of an encoder and a decoder. The encoder embeds the neurons at \emph{both} spatial and temporal levels. The decoder predicts vessel segment flow by utilizing \emph{both} the past flow values and the spatial information of the vessel segments, along with the neuronal activity via the cross-attention mechanism.

\smallskip
\noindent{\bf Transformers\quad} 
The self-attention mechanism introduced by Transformers ~\cite{vaswani2017attention} is based on a trainable associative memory with (key, value) vector pairs, where a query vector $q \in \mathbb{R}^d$ is matched against a set of $k$ key vectors using scaled inner products, as follows
\begin{equation}
\begin{aligned}
\label{transformer_att}
A(Q,K,V)=\text{Softmax}\bigg(\frac{QK^{T}}{\sqrt{d}}\bigg)V,
\end{aligned}
\end{equation}
where $Q \in \mathbb{R}^{N \times d}$, $K \in \mathbb{R}^{k \times d}$ and $V \in \mathbb{R}^{k \times d}$ represent the packed $N$ queries, $k$ keys and values tensors respectively.
Keys, queries and values are obtained using linear transformations of the sequence's elements.
A multi-head self-attention layer is defined by extending the self-attention mechanism using $h$ attention \emph{heads}, i.e. $h$ self-attention functions applied to the input, reprojected to values via a $dh \times D$ linear layer.

\smallskip
\noindent{\bf Neuronal Encoding\quad}
To obtain the initial Spatio-Temporal Encoding, for the prediction at time $t$, we project each neuron to a high $d$ dimensional embedding $\phi^{s}_{t} \in \mathbb{R}^{t_{s}\times n\times d}$ 
by modulating it with its spike value such that $\phi^{s}_{t} = S_{t}\odot ({1}_{d}W^{T})$,
where \mbox{$W \in \mathbb{R}^{d}$} denotes the neuronal encoding. 
The embedding is modulated by the magnitude of the spike, such that higher neuronal activities are projected farther in the embedding space. 

The \emph{temporal encoding} is defined using sinusoidal encoding \cite{vaswani2017attention} applied on $\phi$ and augmented with a learnable embedding such that 
$\phi^{s}_{t}\leftarrow \phi^{s}_{t}+p_{t}\cdot \tilde{p}$ where $p_{t}$ and $\tilde{p}$ represent the sinusoidal time encoding and the learned vector, respectively. 
We emphasize the fact that, contrary to traditional transformers, the embedding tensor $\phi_{t}$ has an additional spatial dimension such that the tensor is three-dimensional, enabling both spatial and temporal attention.

In order to incorporate the spatial information of the neurons, we propose to insert \emph{spatial encoding} by importing the pairwise information directly into the self-attention layer. For this, we multiply the distance relation by the similarity tensor as follows 
\begin{equation}
\begin{aligned}
\label{transformer_att_cc}
A^{S}(Q,K,D_{S})=\text{Softmax}\bigg(\frac{QK^{T}}{\sqrt{d}}\bigg)\odot \psi_{{S}}(D_{S}),
\end{aligned}
\end{equation}
with $\odot$ denoting the Hadamard product, and $\psi_{{S}}(D_{S}):\mathbb{R}^{+} \rightarrow \mathbb{R}^{+}$ an element-wise learnable parameterized similarity function.
This way, the similarity function scales the self-attention map according to the distance between the elements (in our case the neurons).

\smallskip
\noindent{\bf Vascular Decoding\quad} 
The spatio-temporal encoding of the vascular data is similar to the embedding performed by the encoder.
The information on each vascular segment is embedded in a high-dimensional vector $\phi^{F}_{t} \in \mathbb{R}^{t_{v}\times m\times d}$ to be further projected by the temporal encoding.
The spatial geometric information is incorporated via the pairwise vascular segments' distance matrix $D_{F}$ via the decoder's self-attention module $A^{F}$.

The most important element of the decoder is the cross-attention module, which incorporates neuronal information for vascular prediction.
Given the final neuronal embeddings $\phi^{s}_{t}$, the cross-attention module performs cross-analysis of the neuronal embeddings such that
\begin{equation}
\begin{aligned}
\label{transformer_att_cc2}
A^{SF}(Q_{F},K_{S},D_{S})=\text{Softmax}\bigg(\frac{Q_{F}K_{S}^{T}}{\sqrt{d}}\bigg)\odot \psi_{{SF}}(D_{SF}),
\end{aligned}
\end{equation}
where $Q_{F}$ and $K_{S}$ represent the affine transform of $\phi^{F}_{t}$ and $\phi^{s}_{t}$, respectively.
Here also, the (non-square) cross-attention map is modulated by the neuron-vessel distance matrix $D_{SF}$.

The spatio-temporal map is of dimensions $A^{SF}\in\mathbb{R}^{t_{v}\times t_{s}\times h\times m\times n}$ where $h$ denotes the number of attention heads.
Thus, we perform aggregation by averaging over the neuronal time dimension, in order to remain \emph{invariant} to the temporal neuronal embedding and to gather all past neuronal influence on blood flow rates.
This way, one can observe that the proposed method is not limited to any spatial or time constraint.
The model can be deployed in different spatio-temporal settings at test time, thanks to both the geometric spatial encoding and the Transformer's sequential processing ability. 
Finally, the output module reprojects the last time vessel embedding into the prediction space.

\begin{figure}[t]
\centering
\includegraphics[width=0.6\textwidth]{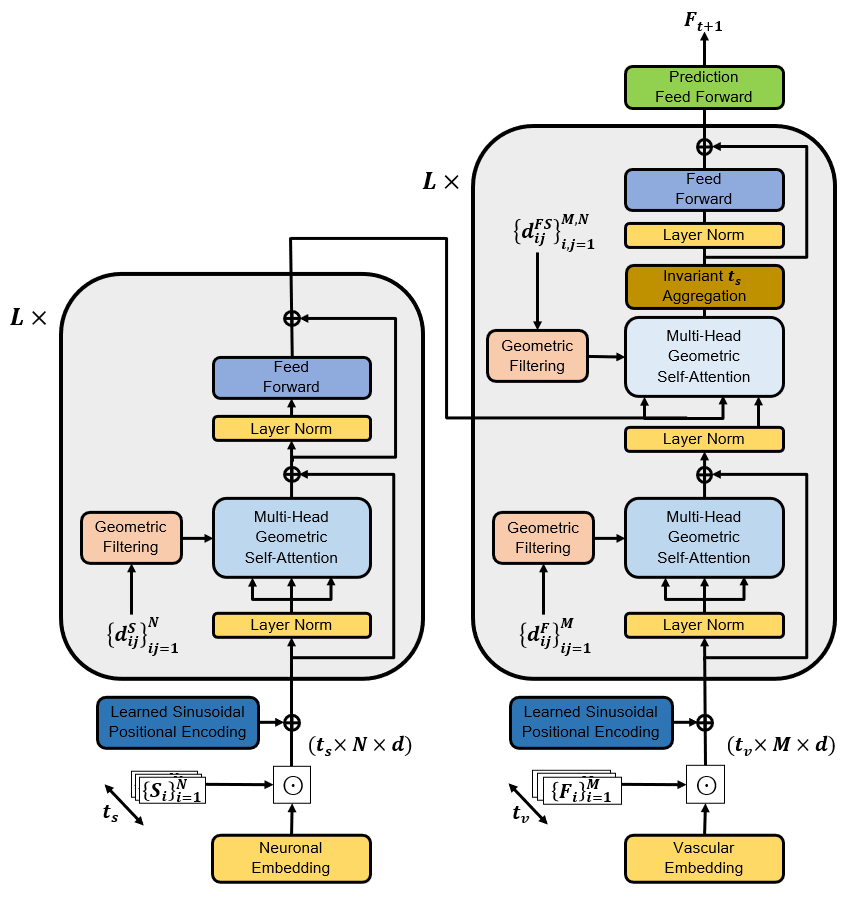}
\caption{Illustration of the proposed HRF Transformer architecture.
The main differences from the traditional Transformers are the Geometric self-attention modules and the unified spatiotemporal analysis induced by the time aggregation module.}
\label{fig:hrft_arch}
\end{figure}

\smallskip
\noindent{\bf Architecture and Training\quad} The initial encoding defines the model embedding dimension $d=64$.
The encoder and the decoder are defined as the concatenation of $L=3$ layers, each composed of self-attention and feed-forward layers interleaved with normalization layers.
The decoder also contains $N$ additional cross-attention modules.
The output layer is defined by a fully connected layer that projects the last vascular time embedding into the objective dimension $m$.
An illustration of the model is given in Figure \ref{fig:hrft_arch}.

The dimension of the feed-forward network is four times that of the embedding \cite{vaswani2017attention}. It is composed of GEGLU layers \cite{shazeer2020glu}, with layer normalization set to the pre-layer norm setting, as in \cite{opennmt,xiong2020layer}.
We use an eight-head self-attention module in all experiments. 
The geometric filtering first augments the distance using Fourier features \cite{tancik2020fourier} and the module is a fully connected neural network with two 50-dimensional hidden layers and GELU non-linearities, expanded to all the heads of the self-attention module.
We provide the module with the element-wise inverse of the distance matrix instead of the regular Euclidean matrix, both in order to reduce the dynamic range and since closer elements may have a higher impact.  

 The training objective is the Mean Squared Error loss 
\begin{equation}
\begin{aligned}
\label{eq:loss}
\mathcal{L} = \mathbb{E}_{t}\bigg(\sum_{j}^{m}\|f_{\theta}( S_{t},F_{t}, D_{S}, D_{F}, D_{SF})-F_{t+1}\|^2 \bigg)
\end{aligned}
\end{equation}

The Adam optimizer \cite{kingma2014adam} is used with 32 samples per minibatch, for 300 epochs.
We initialized the learning rate to $5\cdot10^{-5}$ coupled with a cosine decay scheduler down to $1\cdot 10^{-6}$ at the end of the training. 
The dataset of the first data collection session has been split by $85\%,7.5\%$ and $7.5\%$ for the training, validation, and testing set, respectively.
Training time is approximately 20 hours for time windows $t_{s}=t_{v}=10$, on an NVIDIA RTX A600.
Testing time is approximately 0.25ms per sample.
%

\section{Experiments}
We compare the proposed method, dubbed \textbf{H}emodynamic \textbf{R}esponse \textbf{F}unction \textbf{T}ransformer (HRFT), with several popular statistical and machine-learning models: (i) naive persistence model, which predicts the previous time step's vascular input, (ii) linear regression, which concatenates all the input (blood flow and neuronal data) from all times stamps before performing the regression,  and (iii) a Recurrent Neural Network composed of two stacked GRU~\cite{cho2014properties} layers. 
All the methods are experimented with using the same inputs and the models have similar capacity ($\sim$0.5M parameters).

In order to understand the impact of neuronal information, we also compare our method with the HRFT encoder only applied to the vascular input, referred to as HRFT-S. The only difference between this model and the full HRFT is the cross-attention module. If the neuronal input is irrelevant or the link is too weak to improve the prediction, HRFT is expected not to outperform HRFT-S, which makes the comparison pertinent.

\begin{table}[t]
\caption{The prediction errors of the various methods on two test sets. The first is obtained in the same session for which the training samples were collected (top half of the table, and the second in a separate session 30 minutes later (bottom half).}
\label{tab:results}
\begin{center}
\begin{tabular}{l@{~~~~~~}cc@{~~~~~~}cc@{~~~~~~}cc} 
 \toprule
     &  \multicolumn{2}{c}{ 6 Hz} & \multicolumn{2}{c}{15 Hz} & \multicolumn{2}{c}{30.03 Hz} \\
    \cmidrule(lr){2-3}
    \cmidrule(lr){4-5}
    \cmidrule(lr){6-7}
       Method & MSE & NRMSE  & MSE & NRMSE  & MSE & NRMSE \\
 \midrule
 Persistence         & 24.38 & 0.220 & 12.93 & 0.160 & 6.923 & 0.115  \\ 
 Linear              & 13.65 & 0.166  & 9.660 & 0.139  & 5.911 & 0.107   \\
 RNN                 & 13.78 & 0.168  & 11.38 & 0.157  & 10.31 & 0.147   \\
 HRFT-S      & 13.34 & 0.165  & 9.426 & 0.138 & \textbf{5.782} & 0.110       \\ 
 HRFT                & \textbf{13.00} & \textbf{0.162}  & \textbf{9.370} & \textbf{0.137}  & 5.783 & \textbf{0.106}  \\
  \midrule
Persistence         & 23.11 & 0.221 & 11.97 & 0.160  & 7.662 & 0.125\\ 
 Linear              & 17.01 & 0.192  & 10.26 & 0.147  & 6.002 & 0.111  \\
 RNN                 & 15.193 & 0.182  & 13.04 & 0.172 &  11.87 & 0.162   \\
 HRFT-S      & 14.63 & 0.176  & 9.820 & 0.147  & 5.914 & \textbf{0.110}       \\ 
 HRFT                & \textbf{14.34} & \textbf{0.173} &  \textbf{9.191} & \textbf{0.143}  & \textbf{5.908} & \textbf{0.110}  \\
 \bottomrule

 \end{tabular}
\end{center}
\end{table}

We present both MSE and Normalized Root MSE. 
Because of computational constraints, we randomly subsample 55 vessel segments among the 425.
We trained the model with temporal windows of size $t_{s}=t_{v}=10$, equivalent to 300 ms according to the original data acquisition's $30.03$Hz sampling rate. 

In addition to the original sampling rate, we also present results for prediction based on lower frequencies, in order to check the ability of the models to capture longer-range dependencies. 
We note that the error in these cases is expected to be larger, since the time gap between the last measurement and the required prediction is larger.

In order to check the generalization abilities of the methods, we test the trained models on a second dataset obtained 30 minutes after the sampling of the original dataset (that includes training, validation, and the first test set).  

The results are presented in Tab.~\ref{tab:results}. As can be seen, the HRFT method outperforms all baselines, including the HRFT-S variant, for 6Hz and 15Hz. 
At the original sampling rate, the performance of HRFT and HRFT-S is similar and better than the baselines. 
This is expected since at this frame rate the history of $t_v=10$ we employ spans only 300ms, which is at the limit of the shortest known neurovascular response reported in the literature~\cite{Uluda2017}. 
It is reassuring that error levels for HRFT remain similar for samples taken 30min after the training set (and the first test set) were collected.

\begin{figure}[t]
    \centering
    \begin{tabular}{ccc}
        \includegraphics[width=0.33\linewidth]{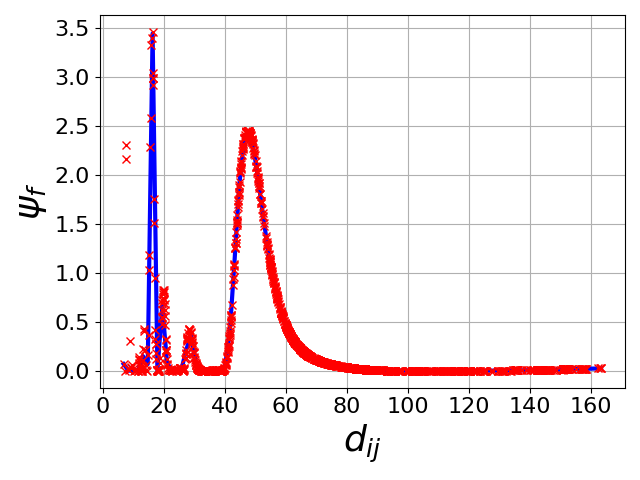}
         &      \includegraphics[width=0.33\linewidth]{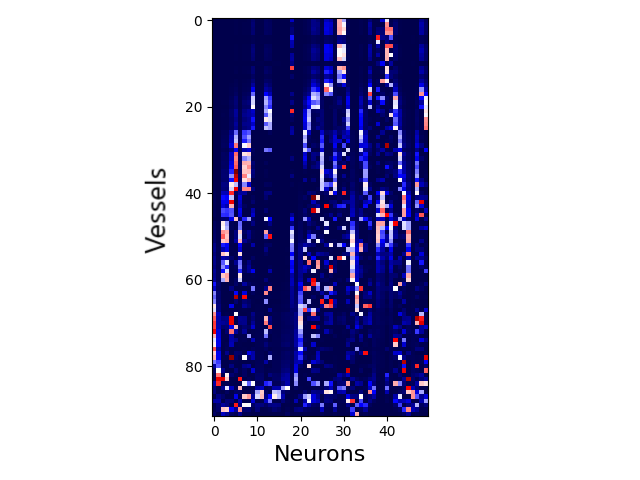}
         &
         \includegraphics[width=0.33\linewidth]{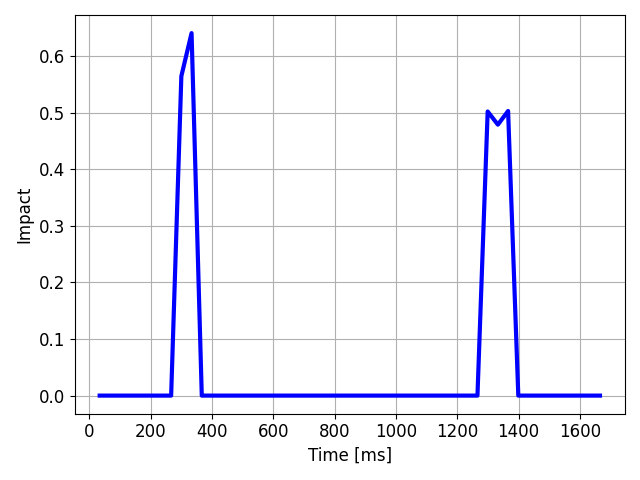}\\
         (a) & (b) & (c)\\
    \end{tabular}
    \caption{(a) The learned function $\psi_{SF}$ for the 1st cross-attention layer of the transformer. 
    (b) The magnitude of the self-attention map between every neuron and every vessel at this layer. 
    (c) The impact of the neurons on the vessels (saturated at 90\%) for each shift in time as obtained by marginalizing over all 2nd session test samples in the 30.02Hz dataset.
    More visualizations of the datasets and the learned features are provided in the Appendix.}
    \label{fig:impact}
\end{figure}

To gain insights into the HRF, we examine the HRFT model. The learned distance function $\psi_{SF}$ of the 1st cross attention layer is depicted in  Fig.~\ref{fig:impact}(a)  (other layers are  similar). The plot shows the learned function in blue and the actual samples in red. Evidently, this prior on the attention is monotonically decreasing with the distance between the neuron and the blood vessel. Panel (b) shows the cross-attention in the same layer. We note that some neurons have little influence, and the rest of the attention is scattered relatively uniformly. Panel (c) considers the derivative of the prediction vector $F_{t+1}$ by each of the neuron data, summed over all test samples of the 2nd session at 6Hz, and all neurons and vessels. There are two negative peaks (contractions) that occur at 333ms and 1333ms, which is remarkably consistent with current knowledge~\cite {Uluda2017}. There is also a dilation effect at 666ms. The 15hz data with $t_v=10$ captures shifts of 0-700ms and the 300ms peak is clearly visible in that model as well.

\section{Conclusions and future work}

We present the first local HRF model. While for the baseline methods, the performance is at the same level with and without neuronal data (omitted from the tables), the transformer we present supports an improved prediction capability using neuronal firing rates (ablation) and also gives rise to interesting insights regarding the behavior of the hemodynamic response function.

\smallskip
\noindent{\bf Limitations\quad} Our main goal is to verify the ability to model HRF by showing that using neuronal data helps predict blood flow beyond the history of the latter.
The next challenge is to scale the model in order to be able to model more vessels (without subsampling) and longer historical sequences (larger $t_v,t_s$). With transformers being used for very long sequences, this is a limitation of our resources and not of our method. 

\section*{Acknowledgements}
This project has received funding from the ISRAEL SCIENCE FOUNDATION
(grant No. 2923/20) within the Israel Precision Medicine Partnership program. It was also supported by a grant from the Tel Aviv University Center for AI and Data Science (TAD).
The contribution of the first author is part of a PhD thesis research conducted at Tel Aviv University.

 \bibliographystyle{splncs04}
 \bibliography{mybibliography}

\begin{thebibliography}{10}
\providecommand{\url}[1]{\texttt{#1}}
\providecommand{\urlprefix}{URL }
\providecommand{\doi}[1]{https://doi.org/#1}

\bibitem{asl2017tracking}
Asl, M.E., Koohbanani, N.A., Frangi, A.F., Gooya, A.: Tracking and diameter
  estimation of retinal vessels using gaussian process and radon transform.
  Journal of Medical Imaging  \textbf{4}(3),  034006 (2017)

\bibitem{buxton2023thermodynamic}
Buxton, R.B.: Thermodynamic limitations on brain oxygen metabolism:
  physiological implications. bioRxiv pp. 2023--01 (2023)

\bibitem{carion2020end}
Carion, N., Massa, F., Synnaeve, G., Usunier, N., Kirillov, A., Zagoruyko, S.:
  End-to-end object detection with transformers. arXiv preprint
  arXiv:2005.12872  (2020)

\bibitem{cho2014properties}
Cho, K., van Merri{\"e}nboer, B., Bahdanau, D., Bengio, Y.: On the properties
  of neural machine translation: Encoder--decoder approaches. In: Proceedings
  of SSST-8, Eighth Workshop on Syntax, Semantics and Structure in Statistical
  Translation. pp. 103--111 (2014)

\bibitem{Drew2019}
Drew, P.J.: Vascular and neural basis of the bold signal. Current Opinion in
  Neurobiology  \textbf{58},  61--69 (10 2019).
  \doi{10.1016/J.CONB.2019.06.004}

\bibitem{drew2010rapid}
Drew, P.J., Blinder, P., Cauwenberghs, G., Shih, A.Y., Kleinfeld, D.: Rapid
  determination of particle velocity from space-time images using the radon
  transform. Journal of Computational Neuroscience  \textbf{29}(1),  5--11
  (2010)

\bibitem{fazlollahi2014efficient}
Fazlollahi, A., Meriaudeau, F., Villemagne, V.L., Rowe, C.C., Yates, P.,
  Salvado, O., Bourgeat, P.: Efficient machine learning framework for
  computer-aided detection of cerebral microbleeds using the radon transform.
  In: 2014 IEEE 11th International Symposium on Biomedical Imaging (ISBI). pp.
  113--116. IEEE (2014)

\bibitem{gao2014determination}
Gao, Y.R., Drew, P.J.: Determination of vessel cross-sectional area by
  thresholding in radon space. Journal of Cerebral Blood Flow \& Metabolism
  \textbf{34}(7),  1180--1187 (2014)

\bibitem{Kim2012}
Kim, S.G., Ogawa, S.: Biophysical and physiological origins of blood
  oxygenation level-dependent fmri signals. Journal of cerebral blood flow and
  metabolism : official journal of the International Society of Cerebral Blood
  Flow and Metabolism  \textbf{32},  1188--206 (7 2012).
  \doi{10.1038/jcbfm.2012.23}

\bibitem{kingma2014adam}
Kingma, D., Ba, J.: Adam: A method for stochastic optimization. arXiv  (2014)

\bibitem{opennmt}
Klein, G., Kim, Y., et~al.: Open-source toolkit for neural machine translation.
  In: ACL (2017)

\bibitem{levy2021communication}
Levy, W.B., Calvert, V.G.: Communication consumes 35 times more energy than
  computation in the human cortex, but both costs are needed to predict synapse
  number. Proceedings of the National Academy of Sciences  \textbf{118}(18),
  e2008173118 (2021)

\bibitem{lewis2019bart}
Lewis, M., Liu, Y., Goyal, N., Ghazvininejad, M., Mohamed, A., Levy, O.,
  Stoyanov, V., Zettlemoyer, L.: Bart: Denoising sequence-to-sequence
  pre-training for natural language generation, translation, and comprehension.
  arXiv preprint arXiv:1910.13461  (2019)

\bibitem{li2019visualbert}
Li, L.H., Yatskar, M., Yin, D., Hsieh, C.J., Chang, K.W.: Visualbert: A simple
  and performant baseline for vision and language. arXiv preprint
  arXiv:1908.03557  (2019)

\bibitem{li2020oscar}
Li, X., Yin, X., Li, C., Zhang, P., Hu, X., Zhang, L., Wang, L., Hu, H., Dong,
  L., Wei, F., et~al.: Oscar: Object-semantics aligned pre-training for
  vision-language tasks. In: European Conference on Computer Vision. pp.
  121--137. Springer (2020)

\bibitem{Logothetis2008a}
Logothetis, N.K.: What we can do and what we cannot do with fmri. Nature
  \textbf{453},  869--78 (6 2008). \doi{10.1038/nature06976}

\bibitem{lu2019vilbert}
Lu, J., Batra, D., Parikh, D., Lee, S.: Vilbert: Pretraining task-agnostic
  visiolinguistic representations for vision-and-language tasks. In: Advances
  in Neural Information Processing Systems. pp. 13--23 (2019)

\bibitem{mookiah2020review}
Mookiah, M.R.K., Hogg, S., MacGillivray, T.J., Prathiba, V., Pradeepa, R.,
  Mohan, V., Anjana, R.M., Doney, A.S., Palmer, C.N., Trucco, E.: A review of
  machine learning methods for retinal blood vessel segmentation and
  artery/vein classification. Medical Image Analysis p. 101905 (2020)

\bibitem{paul2021local}
Paul, M., Danelljan, M., Van~Gool, L., Timofte, R.: Local memory attention for
  fast video semantic segmentation. arXiv preprint arXiv:2101.01715  (2021)

\bibitem{pourreza2008radon}
Pourreza, R., Banaee, T., Pourreza, H., Kakhki, R.D.: A radon transform based
  approach for extraction of blood vessels in conjunctival images. In: Mexican
  International Conference on Artificial Intelligence. pp. 948--956. Springer
  (2008)

\bibitem{shazeer2020glu}
Shazeer, N.: Glu variants improve transformer. arXiv:2002.05202  (2020)

\bibitem{tan2019lxmert}
Tan, H., Bansal, M.: Lxmert: Learning cross-modality encoder representations
  from transformers. arXiv preprint arXiv:1908.07490  (2019)

\bibitem{tancik2020fourier}
Tancik, M., Srinivasan, P., Mildenhall, B., Fridovich-Keil, S., Raghavan, N.,
  Singhal, U., Ramamoorthi, R., Barron, J., Ng, R.: Fourier features let
  networks learn high frequency functions in low dimensional domains. Advances
  in Neural Information Processing Systems  \textbf{33},  7537--7547 (2020)

\bibitem{tavakoli2011radon}
Tavakoli, M., Mehdizadeh, A., Pourreza, R., Pourreza, H.R., Banaee, T., Toosi,
  M.B.: Radon transform technique for linear structures detection: application
  to vessel detection in fluorescein angiography fundus images. In: 2011 IEEE
  Nuclear Science Symposium Conference Record. pp. 3051--3056. IEEE (2011)

\bibitem{Uluda2017}
Uludağ, K., Blinder, P.: Linking brain vascular physiology to hemodynamic
  response in ultra-high field mri. NeuroImage  \textbf{168},  279--295 (3
  2018). \doi{10.1016/j.neuroimage.2017.02.063}

\bibitem{vaswani2017attention}
Vaswani, A., Shazeer, N., Parmar, N., et~al.: Attention is all you need. In:
  NeurIPS (2017)

\bibitem{wang2020max}
Wang, H., Zhu, Y., Adam, H., Yuille, A., Chen, L.C.: Max-deeplab: End-to-end
  panoptic segmentation with mask transformers. arXiv preprint arXiv:2012.00759
   (2020)

\bibitem{wang2020end}
Wang, Y., Xu, Z., Wang, X., Shen, C., Cheng, B., Shen, H., Xia, H.: End-to-end
  video instance segmentation with transformers. arXiv preprint
  arXiv:2011.14503  (2020)

\bibitem{xiong2020layer}
Xiong, R., et~al.: On layer normalization in the transformer architecture.
  arXiv:2002.04745  (2020)

\bibitem{zhu2021deformable}
Zhu, X., Su, W., Lu, L., Li, B., Wang, X., Dai, J.: Deformable {DETR}:
  Deformable transformers for end-to-end object detection. In: International
  Conference on Learning Representations (2021),
  \url{https://openreview.net/forum?id=gZ9hCDWe6ke}

\end{thebibliography}
\newpage
\appendix
\section{Additional visualizations}
We provide several visualizations of the data and the induced learned feature maps.
The data is selected such that we remove vessels with diameters less than the data acquisition resolution.
\subsection{Data Visualization}
\begin{figure}[h!]
    \centering
        \includegraphics[width=0.49\linewidth]{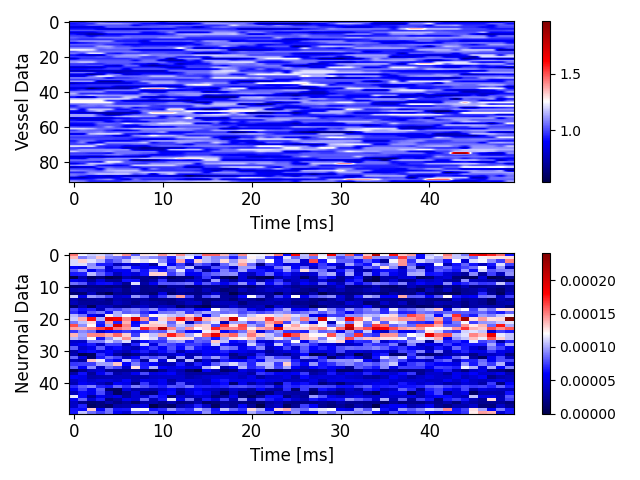}
          \includegraphics[width=0.49\linewidth]{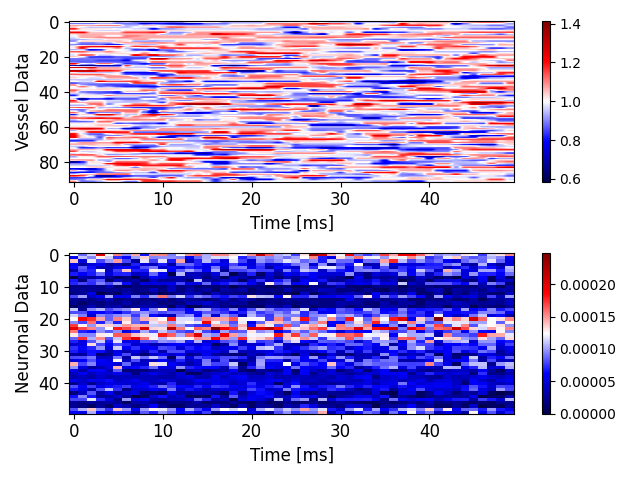}
         \includegraphics[width=0.49\linewidth]{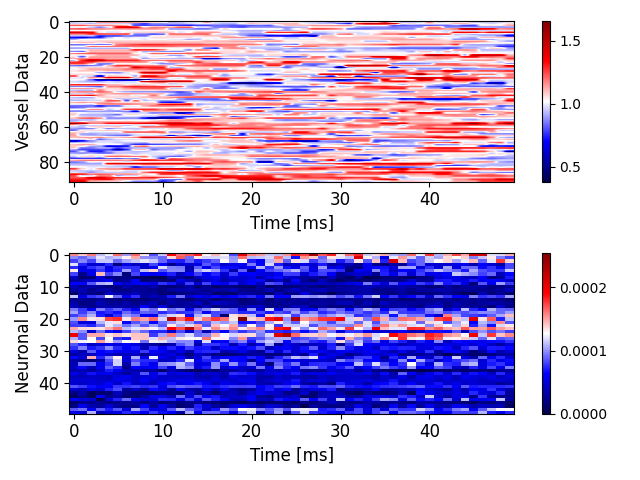}
         \includegraphics[width=0.49\linewidth]{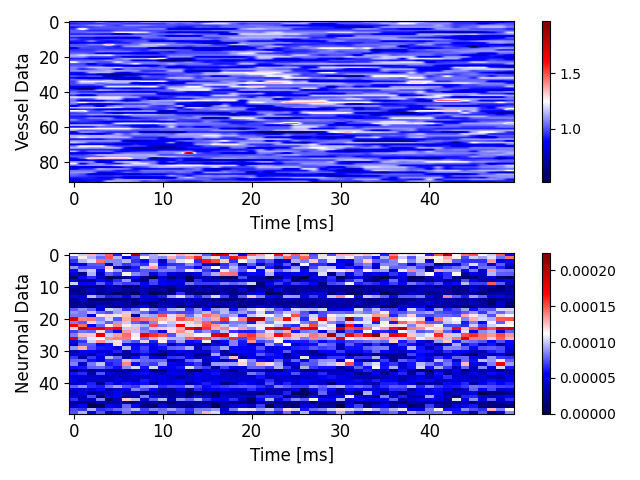}
    \caption{Different samples of the original 30.02Hz sampled vascular and neuronal data.}
    \label{appendix:data_samples}
\end{figure}

In Figure \ref{appendix:data_samples} we provide the visualization of a few samples from the dataset sampled at 30.02Hz.
\subsection{Data Geometry Visualization}
\begin{figure}[h!]
    \centering
        \includegraphics[width=0.32\linewidth]{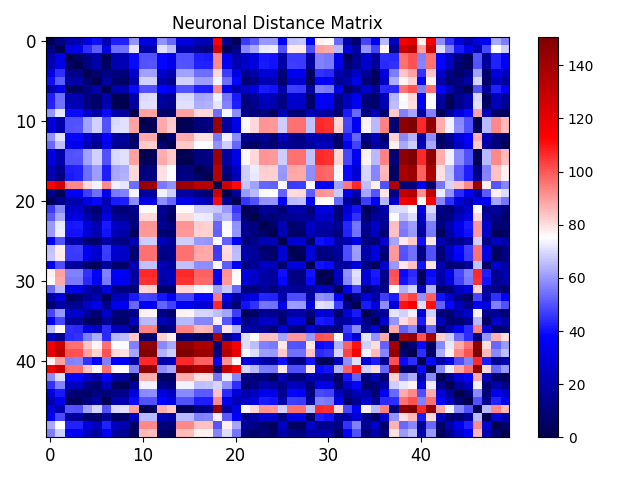}
          \includegraphics[width=0.32\linewidth]{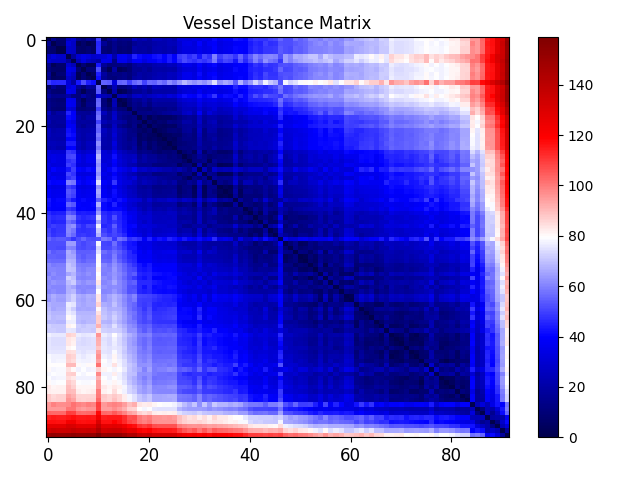}
         \includegraphics[width=0.32\linewidth]{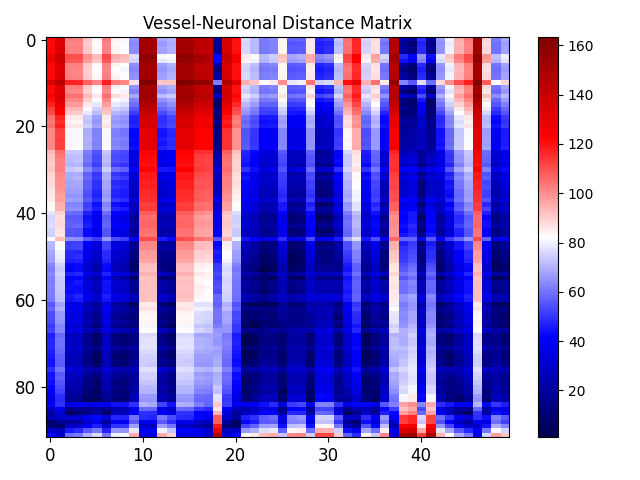}
         
    \caption{Distances matrices.}
    \label{appendix:dist_mat}
\end{figure}

In Figure \ref{appendix:dist_mat} we provide the visualization of the different pairwise distance matrices used in the model.
\subsection{Geometric Filters Visualization}
\begin{figure}[h!]
    \centering
        \includegraphics[width=0.32\linewidth]{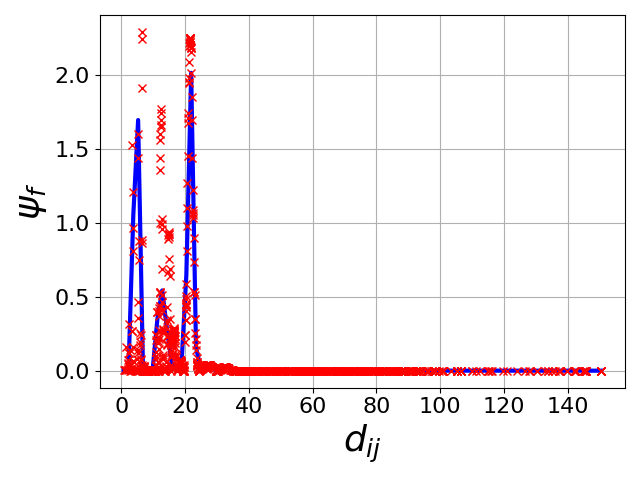}
          \includegraphics[width=0.32\linewidth]{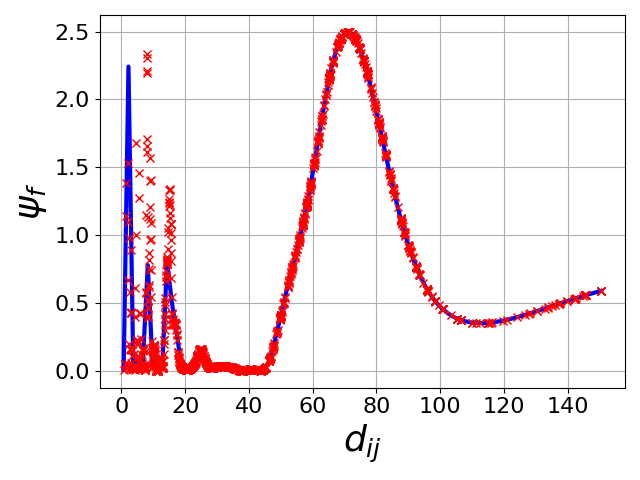}
         \includegraphics[width=0.32\linewidth]{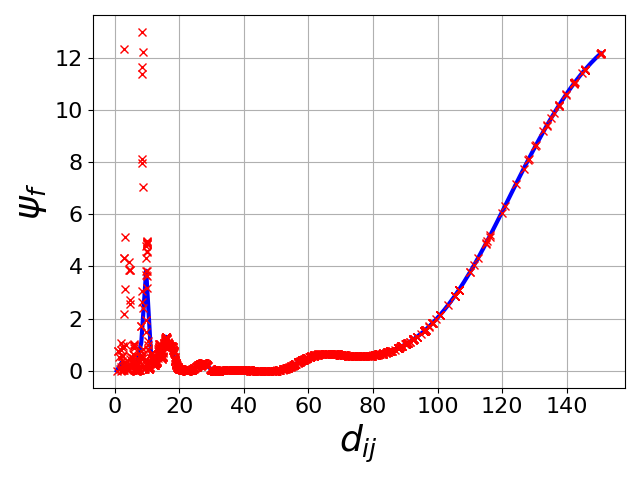}
         \\
         \includegraphics[width=0.32\linewidth]{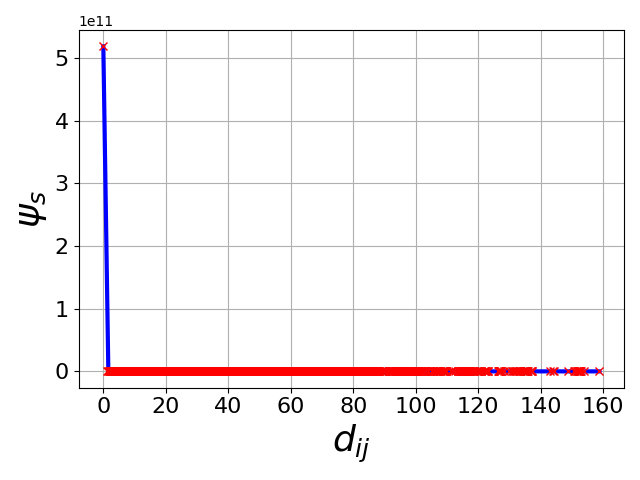}
          \includegraphics[width=0.32\linewidth]{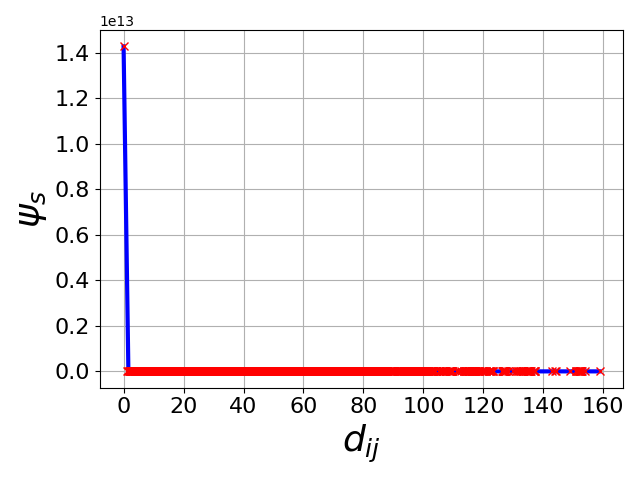}
         \includegraphics[width=0.32\linewidth]{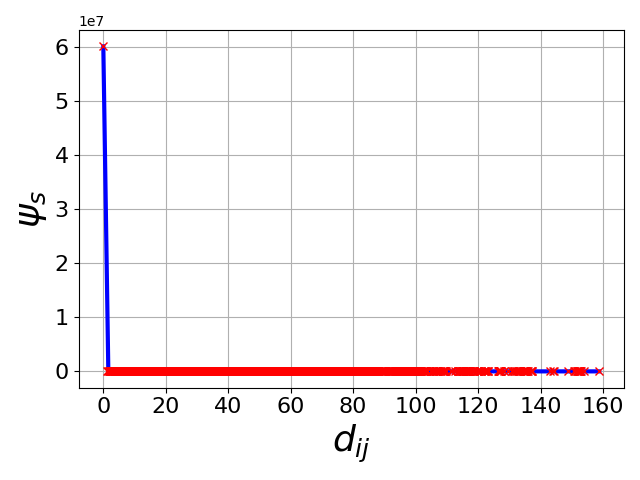}
         \\
         \includegraphics[width=0.32\linewidth]{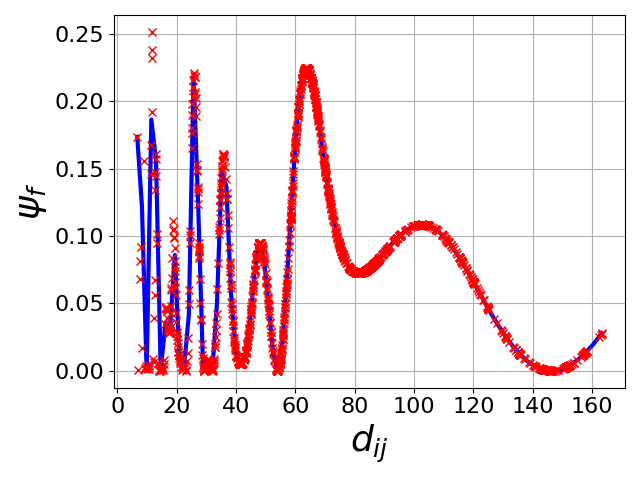}
          \includegraphics[width=0.32\linewidth]{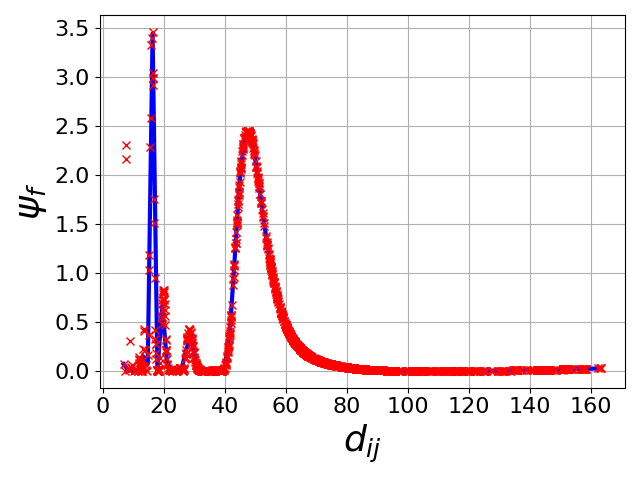}
         \includegraphics[width=0.32\linewidth]{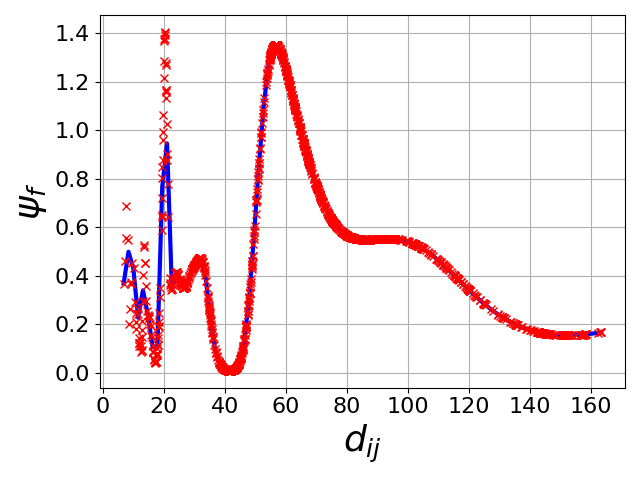}
    \caption{Learned Geometric filters. First row: Neuronal filtering. Second row: vascular filtering. Third row: Neuro-vascular filtering.
    We can observe the vessels' interactions are impactful only between extremely close vessels.}
    \label{appendix:geo_filter}
\end{figure}

In Figure \ref{appendix:geo_filter} we provide the visualization of the learned distance filters.

\subsection{Self-Attention Map Visualization}
\begin{figure}[h!]
    \centering
        \includegraphics[width=0.32\linewidth]{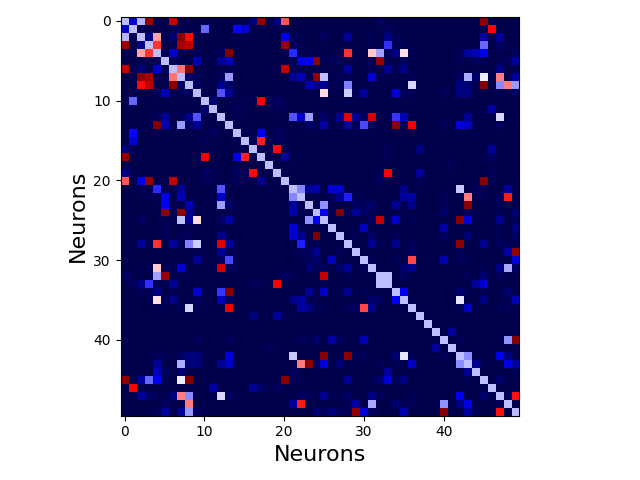}
          \includegraphics[width=0.32\linewidth]{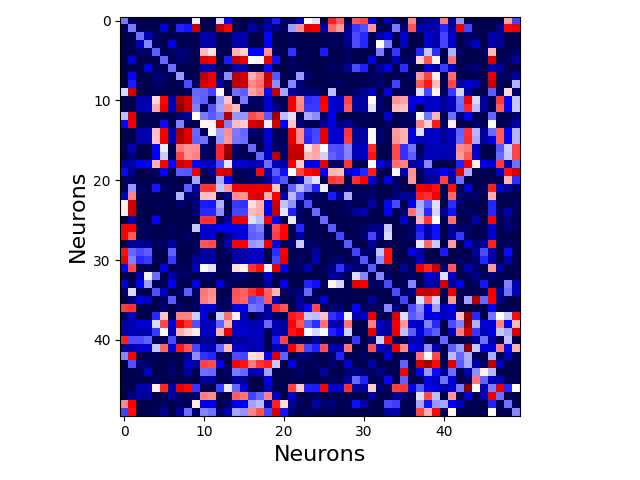}
         \includegraphics[width=0.32\linewidth]{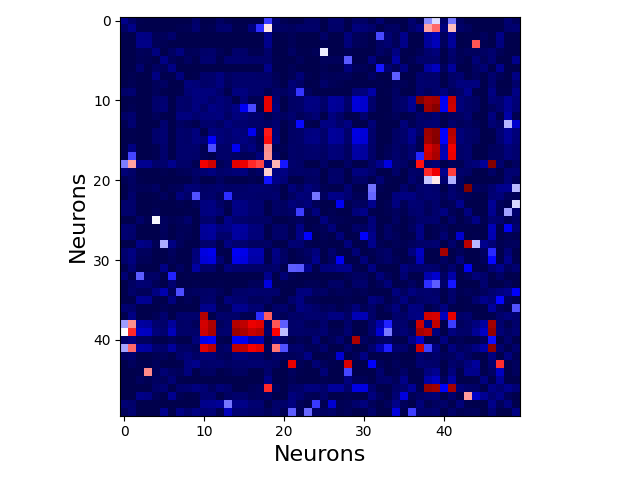}
         \\
         \includegraphics[width=0.32\linewidth]{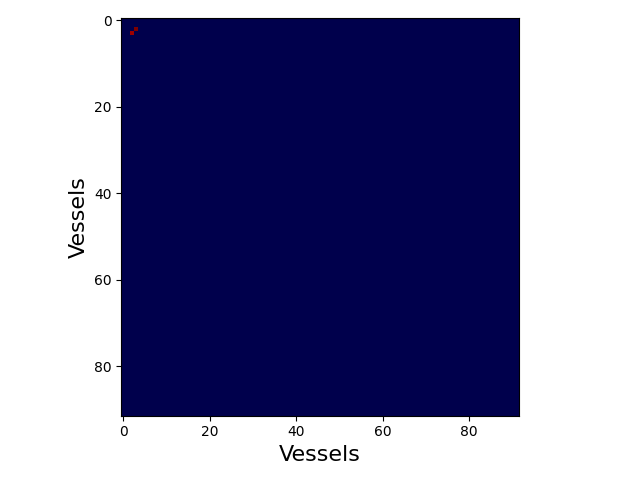}
          \includegraphics[width=0.32\linewidth]{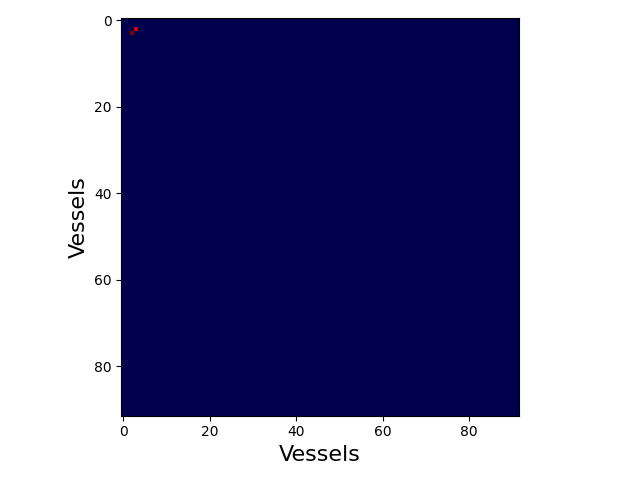}
         \includegraphics[width=0.32\linewidth]{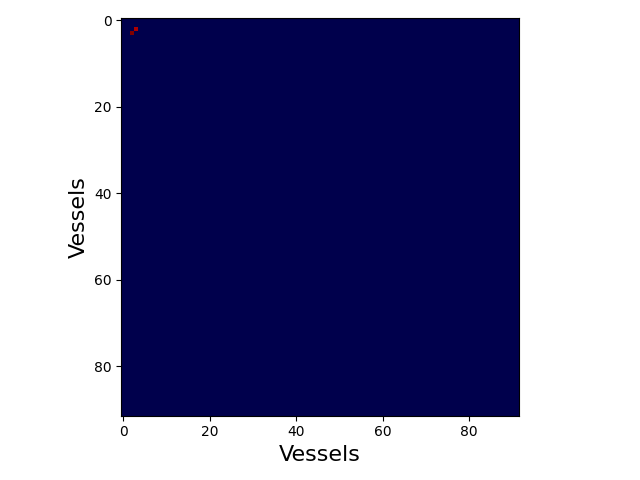}
         \\
         \includegraphics[width=0.32\linewidth]{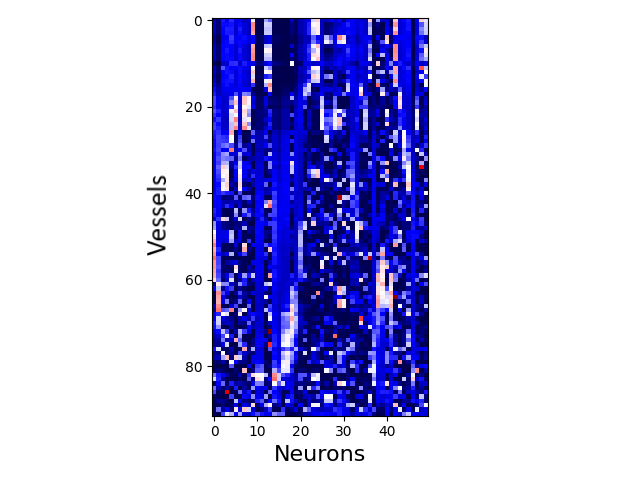}
          \includegraphics[width=0.32\linewidth]{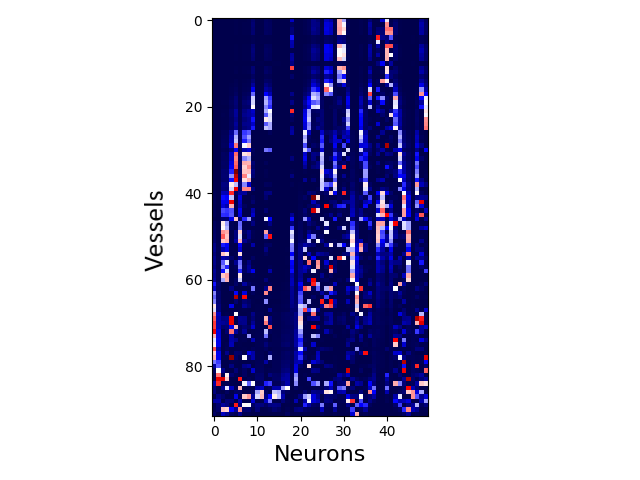}
         \includegraphics[width=0.32\linewidth]{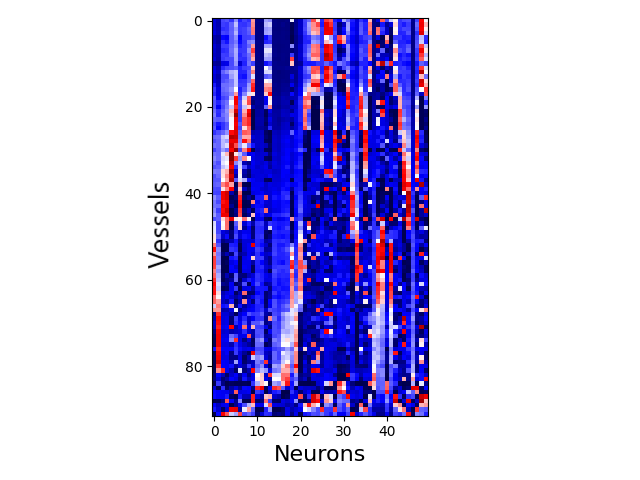}
    \caption{Illustration of the self-attention maps at different layers of the model.
    The vascular self-attention is nearly constant because of the learned vascular filters.}
        \label{appendix:sa_maps}
    \end{figure}

In Figure \ref{appendix:sa_maps} we provide typical visualization of the self-attention maps at different layers of the network.
\subsection{Neuronal Influence}
\begin{figure}[h!]
    \centering
        \includegraphics[width=0.9\linewidth]{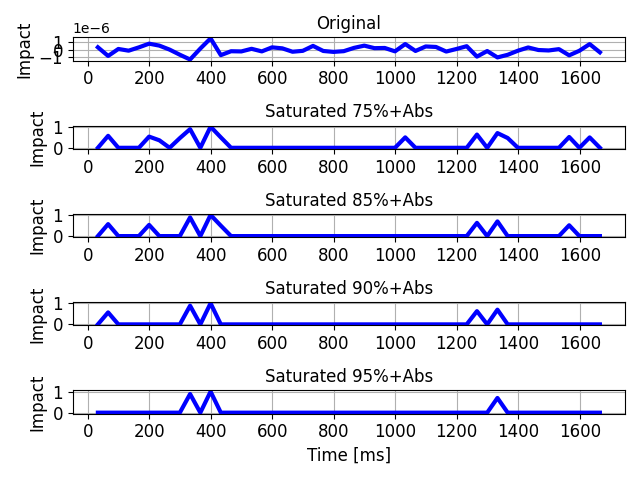}\\
        (a) \\
        \includegraphics[width=0.9\linewidth]{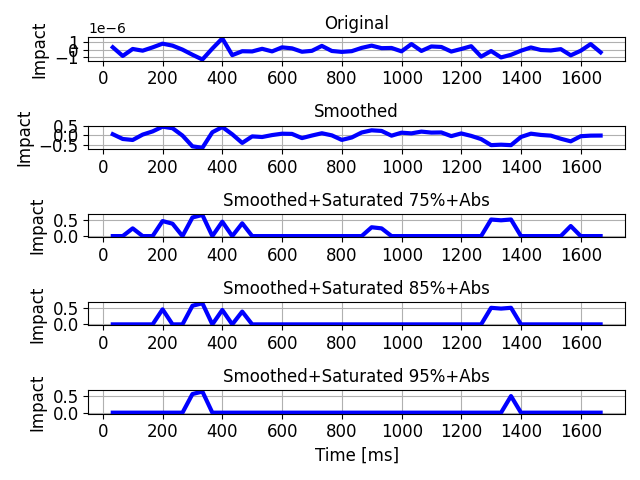}\\
        (b)
            
    \caption{Illustration of the original neuronal gradient data and its saturation in order to obtain the neuronal impact consistent with the literature. (a): non-smoothed signal and different saturation percentiles. (b): smoothed (Savitzky-Golay filter) signal and different saturation percentiles.  }
        \label{appendix:neuronal_impact}
    \end{figure}

In Figure \ref{appendix:neuronal_impact} we provide the visualization of the mean processed  neuronal influence (i.e., gradient) over time.
\end{document}